\documentclass[fleqn,10pt]{wlscirep_arxiv}

\title{A cohesive granular material with tunable elasticity}

\author[1,*]{Arnaud Hemmerle}
\author[1,2]{Matthias Schr\"oter}
\author[1,3,\dag]{Lucas Goehring}
\affil[1]{Max Planck Institute for Dynamics and Self-Organization (MPIDS), G\"ottingen, 37077, Germany}
\affil[2]{Institute for Multiscale Simulation, Friedrich-Alexander-University, Erlangen, 91052, Germany}
\affil[3]{School of Science and Technology, Nottingham Trent University, Clifton Lane, Nottingham, NG11 8NS, UK}

\affil[*]{arnaud.hemmerle@ds.mpg.de} \affil[$\dag$]{lucas.goehring@ntu.ac.uk}

\begin{abstract}
By mixing glass beads with a curable polymer we create a well-defined cohesive granular medium, held together by solidified, and hence elastic, capillary bridges. This material has a geometry similar to a wet packing of beads, but with an additional control over the elasticity of the bonds holding the particles together. We show that its mechanical response can be varied over several orders of magnitude by adjusting the size and stiffness of the bridges, and the size of the particles. We also investigate its mechanism of failure under unconfined uniaxial compression in combination with \textit{in situ} x-ray microtomography. We show that a broad linear-elastic regime ends at a limiting strain of about 8\%, whatever the stiffness of the agglomerate, which corresponds to the beginning of shear failure. The possibility to finely tune the stiffness, size and shape of this simple material makes it an ideal model system for investigations on, for example, fracturing of porous rocks, seismology, or root growth in cohesive porous media.  
\end{abstract}
\begin{document}

\flushbottom
\maketitle

\thispagestyle{empty}

\vspace{-0.8cm}
\section*{Introduction}
The mechanical and fracture properties of cohesive granular materials, where solid but deformable bonds ensure the rigidity of a granular packing, are relevant to a number of questions in powder aggregation \cite{Iveson1996}, the strength of agglomerates \cite{Kendall2001}, soil rheology \cite{Mitchell1976}, rock mechanics \cite{Goodman1989} and geoengineering \cite{Turcotte2014}.   As one example, sandstone is a porous rock widely used in construction, and is composed of grains of sand  held together by a cement of, most often, calcite or silica.  It is also the most relevant material for modelling underground aquifers or hydrocarbon reservoirs.   Since the early work of Dvorkin and his collaborators on cemented aggregates\cite{Dvorkin1991,Dvorkin1994, Dvorkin1999}, several approaches have been developed to overcome the intrinsic complexity of such heterogeneous media.   Most progress has been made in the field of numerical modeling \cite{Jing2003,Potyondy2004,Jiang2013}, while bottom-up experimental approaches using simple systems and general constitutive models are still rare \cite{Delenne2004,Delenne2011,Papamichos1999,Rieser2015}. Generally, one wants to be able to predict or control macroscopic properties of cohesive porous media, such as fluid permeability \cite{Turcotte2014}, fracture response \cite{Affes2012,Birchall1981}, mechanical constitutive relations \cite{Kendall1987,Delenne2004,Delenne2011,Goehring2009}, or oil/water retention \cite{Murison2014}, and to evaluate how these properties depend on the composition of the material, such as grain-scale heterogeneities \cite{Murison2014}, or its elasticity \cite{Holtzman2012,Delenne2009}.  

Here we describe a class of cohesive granular materials with tunable properties, and characterise the mechanical response of these materials to uniaxial compression. The materials are similar to wet granulates, like sand, but where the liquid in the capillary bridges between adjacent grains has been replaced by polydimethylsiloxane (PDMS), a curable elastomer, which is then cross-linked to produce ``solid'' capillary bridges (see Fig.\ref{Beads}.a-c). Before curing, PDMS is a viscous liquid, and perfectly wets the glass beads. By mixing the two phases together (see Methods), a homogeneous material is easily made; the surface tension of the polymer phase naturally attracts it into liquid capillary bridges between particles, and distributes it evenly.  The resulting composite is cohesive and malleable, and may be moulded into any desired shape.   It can then be cured to harden the bridges by cross-linking the polymer, turning it from a paste-like material to a solid cohesive granular assembly. 

For these materials, we focus our attention on conditions of soft bridges, and hard particles.  In particular, in our system the Young's modulus of the polymer bridges, $E_P\sim1$kPa$-1$ MPa, is several orders of magnitude smaller than the Young's modulus of the beads, $E_{b}\sim 60$ GPa.  Thus, we ensure that the mechanical properties of the composite materials are dominated by the deformation of the bridges, and not by the indentation of the beads into each other, such as would be the case with powders \cite{Kendall1987,Johnson1971}, for example. Our materials are, in fact, roughly in between the two limits of wet \cite{Scheel2008, Li2014} or charged \cite{Chen2016} granular materials, where bonds are weak and can re-form after breaking, and weakly sintered \cite{Rice1985, Arato1995} or cemented \cite{Dvorkin1991,Dvorkin1994, Dvorkin1999, Affes2012,Langlois2014} materials, whose bonds are of comparable stiffness with that of the grains, and which are generally brittle.

We will show how one can control the Young's modulus $E$ of such cohesive granular materials by over two orders of magnitude, by changing the modulus $E_p$ and volume fraction $W$ of the polymer phase, and the diameter $D$ of the particles in the granular phase.  A simple model for the micro-contacts, or bridges between particles, is also developed to explain how $E$ should scale with these parameters.   Finally, we report on the ultimate yielding of these materials, through shear failure under large strain, using \textit{in situ} x-ray microtomography.  

\section*{Results}

The mechanical properties of cohesive granular materials depend on their microscopic structure and composition.  Briefly, if the polymer bridges between particles are stiffer, or larger, we found the Young's modulus of the bulk material to be higher, as might be expected. There is also some dependence of the material stiffness on the size of the particles used. 

We prepared material samples with as broad a range of elastic properties as possible, by changing the bead size $D$, polymer fraction $W$ (defined as the total volume of PDMS within a sample, divided by the sample volume), the Young's modulus of the polymer $E_p$, and packing fraction $\phi_b$.  Details of the material preparation are given in the Methods section, at the end of this paper.  Similar materials could also be made with other polymers (such as polyurethane or polybutadiene), extending the range of bulk elastic properties even further than those presented here.  Additionally, the wetting properties of the beads and bridges could be adjusted by judicious choice of ingredients (\textit{e.g.\,}polystyrene beads, surface-treated beads, or more complex heterogeneities\cite{Walther2008,Murison2014} like Janus or patchy particles), as could the permeability of the final product.  

\begin{figure}
\centering
\includegraphics[width=1\linewidth]{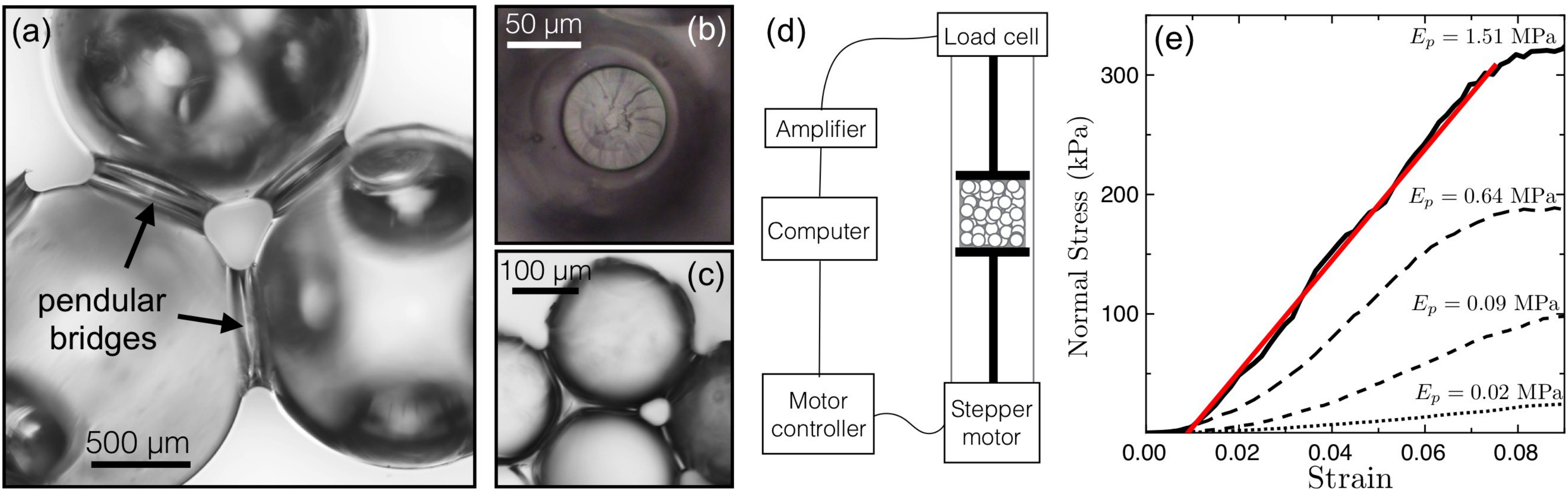}
\caption{(a) Three beads are shown connected by polymer bridges in the form of pendular rings, at a polymer content of $W=2.7$\%.  Other bridges, which have debonded from their neighbours, are also visible, as is (b) a debonded bridge viewed from above.  (c) For higher polymer fractions the pendular rings begin to merge, forming clusters such as the trimer\cite{Scheel2008} shown here for $W = 5.2\%$. (d) Sketch of the uniaxial compression setup. (e) Typical stress-strain curves of the cohesive granular material, as measured for various stiffnesses of the polymer forming the bridges for $W=2.3\%$.  The Young's modulus of a sample, $E$, is found by a least-squares fit to the linear region of its stress-strain curve, as demonstrated by the red line.}
\label{Beads}
\end{figure}

To test the elastic response of our samples, we designed an unconfined uniaxial compression cell, with a geometry also suitable for \textit{in-situ} x-ray microtomography experiments.  The test rig is sketched in Fig.\ref{Beads}.d, and details of its construction, and the test procedures used, are given in the Methods section.   We plot in Fig.\ref{Beads}.e examples of stress-strain curves measured for various $E_p$, at a fixed bead diameter of 210 $\mu$m, and polymer content $W=2.3\%$. All curves show comparable behaviours, and these measurements are typical of all the results presented in this paper.   There is a short non-linear regime, extending to a strain of up to 2\%.  Here, this strain represents a displacement of $\sim$ 100 $\mu$m, or one bead radius. This initial response corresponds to the progressive contact of the upper piston of the load cell with the sample, whose surface is uneven on this scale, rather than any inherent non-linear elasticity.  Once good contact is made between the piston and the sample, further compression shows a well-defined linear regime, from which a Young's modulus $E$ is extracted. Deviation from the linear behaviour is observed after a strain of about 8\% on average, followed by plasticity and shear failure. 

\subsection*{Variation with polymer stiffness}

\begin{figure}
\centering
\includegraphics[width=1\linewidth]{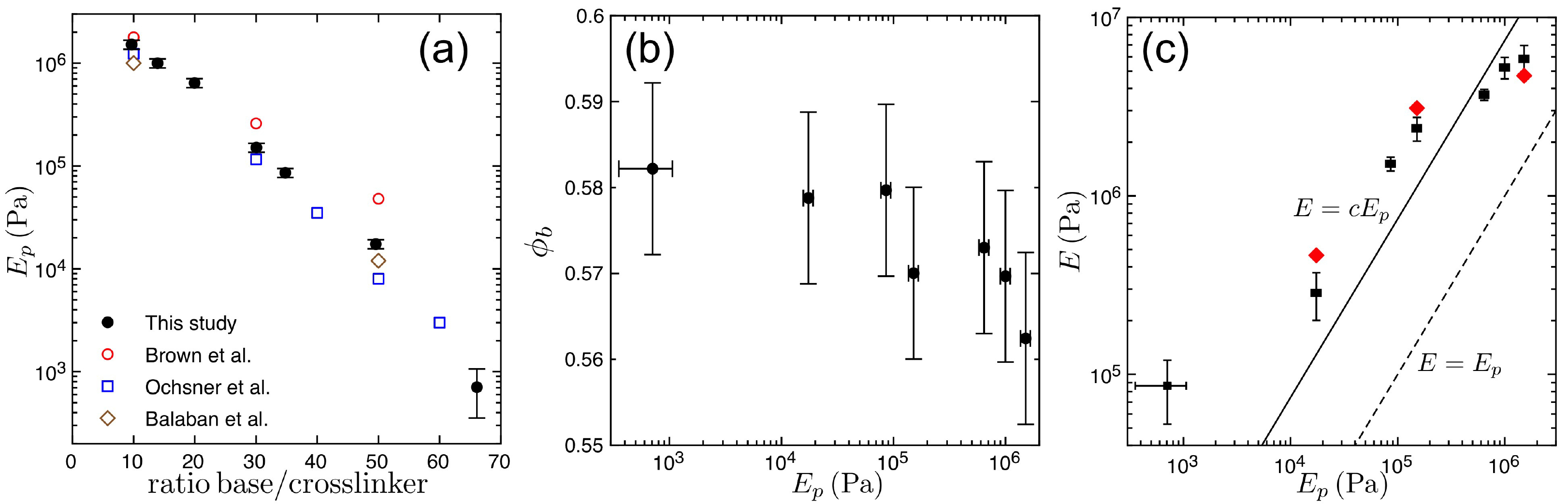}
\caption{(a) The Young's modulus $E_P$ of bulk PDMS depends on the ratio of base to cross-linker used in its preparation.  Our measurements (filled circles) are comparable to those reported in other studies \cite{Brown2005,Ochsner2007,Balaban2001} (open symbols). Error bars show the standard errors of the slopes of linear fits of individual compression tests. (b) For cohesive granular media made using these polymers, the packing fraction of the beads, $\phi_b$, is roughly constant, but may decrease by 1-2$\%$ in the stiffest samples. (c) The Young's modulus of the cohesive granular material $E$ increases with that of the polymer $E_{P}$. For comparison, the dashed line shows $E=E_P$ and the solid line a best-fit linear relation, $E=c E_P$, with $c=7.4$. Error bars for $E$ give the standard deviations of measurements, typically after 4 replicates. Red diamonds: single measurements on larger samples of diameters 15.0 mm x height 7.7 mm were performed to exclude any dependence of $E$ on sample dimensions, as described in the Methods.}
\label{Young}
\end{figure}

Without changing the structure of the cohesive granular material, its elastic response can be adjusted by varying the stiffness of its polymer bridges.    PDMS is a cross-linkable polymer whose Young's modulus can range over several orders of magnitude by changing the ratio of oligomer (base) to cross-linker, prior to curing.  By varying the mass ratio of base to cross-linker from 66:1 to 10:1, we measured Young's moduli of bulk PDMS samples from $E_{P} \sim$ 1 kPa to 1.5 MPa, respectively.  As shown in Fig.\ref{Young}.a, these values are in good agreement with results reported elsewhere \cite{Brown2005,Ochsner2007,Balaban2001}, despite differences in curing times and temperatures.   Outside of this range of mass ratios the stiffness of the PDMS does not vary significantly, as it is already almost completely cross-linked in the harder limit, while for the softest case the PDMS does not fully solidify, and remains tacky to the touch.   Like many other elastomers, PDMS is essentially incompressible, with a Poisson ratio close to 0.5.

Material samples were made by mixing different preparations of PDMS with 210 $\mu$m diameter glass beads.  For these experiments the polymer content was fixed at $W=2.3$\%.   This choice ensures that well-defined capillary bridges connect neighbouring beads, without merging into clusters; as will be described in the next section, we are in the pendular regime\cite{Herminghaus2013}. The bead volume fraction $\phi_b$ was measured in these samples to be, on average, $0.574$, with a standard deviation (representing the reproducibility between samples) of $\pm0.01$.  In general, $\phi_b$ did not vary strongly with $E_{P}$.  However, as Fig.\ref{Young}.b shows, there may be a slight decrease in packing fraction in the hardest samples.

For these conditions, as summarised in Fig.\ref{Young}.c, we found that the Young's modulus $E$ of the cohesive granular material was about an order of magnitude higher than that of the PDMS in its bridges, but very much lower than that of the glass beads.  In other words the elasticity of the composite material increases with, and is controlled by, that of the PDMS, despite the fact that the polymer represents only a small minority phase.  By varying $E_P$  from 1 kPa to 1.5 MPa, we could tune $E$ from 90 kPa to 6 MPa, respectively. Finally we note that at the lowest concentration of cross-linker the PDMS remains somewhat liquid-like even after curing; it still made a rigid cohesive sample, but the properties of this material may be controlled more by capillary forces, rather than the elasticity of the bridges.

\subsection*{Variation with polymer content}

Next, we studied how the size and shape of the polymer bridges affects the elasticity of the cohesive granular material, while keeping all other properties of the polymer and beads constant.  For comparison, it is well-known that the mechanical properties of wet granular materials are linked to their internal liquid morphology \cite{Pietsch1969,Scheel2008,Herminghaus2013}.  In particular, the strength of such materials increases rapidly with liquid content in the pendular regime, \textit{i.e.\,}as long as the fluid remains in individual capillary rings, like those shown in Fig.\ref{Beads}.a.   Beyond some threshold, additional fluid does not create stronger bonds, but instead causes the rings to coalesce into larger structures, like the trimer shown in Fig.\ref{Beads}.c.  Above this so-called pendular-funicular transition, the strength of something like a sandcastle is surprisingly independent of its liquid content \cite{Scheel2008}.   Here we demonstrate similar results for the scaling of the cohesive granular material's Young's modulus, $E$, with polymer content $W$.

\begin{figure}
\centering
\includegraphics[width=0.7\linewidth]{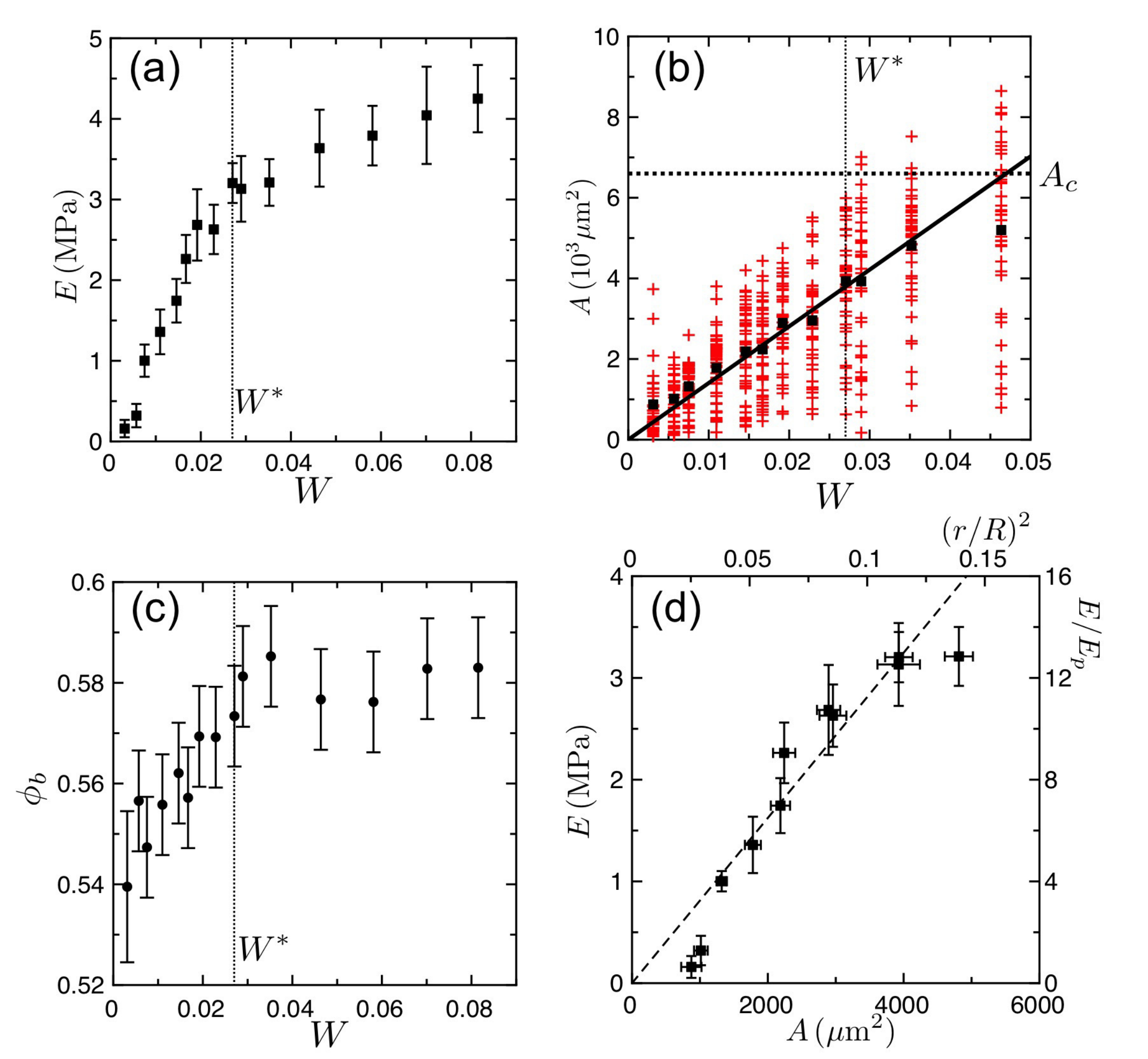}
\caption{(a) The Young's modulus of the cohesive granular material $E$ increases with increasing content of PDMS, $W$. Error bars show standard deviations after 4 replicates. (b) The mean cross-sectional area, $A$, of the pendular polymer bridges also increases with $W$.  Black squares give average values of many individual measurements (red crosses).   The line at $A_c$ corresponds to the predicted maximal size of capillary bridges, before they merge into trimer structures \cite{Scheel2008,Herminghaus2013}. The solid line shows a linear fit to the average values of $A$ for $W\leq3.5\%$. (c) Volume fraction of the beads $\phi_b$ as a function of $W$.  (d) $E$ as a function of the mean bridge cross-sectional area $A$, for $0.3\%\leq W \leq3.5\%$. The dashed line represents a linear relation between $E$ and $A$.  }
\label{PhiPDMS}
\end{figure}

For this set of experiments we prepared a range of samples with different polymer contents, from $W=0.3\%$ to 8.2$\%$, for fixed bead size $D=210 \,\mu$m and polymer modulus $E_{P}=250$ kPa.   The range of $W$ studied allows us to probe the pendular regime, starting from the first observable capillary bridges at low $W$, but also extends to the funicular regime, as we will shortly demonstrate.  As shown in Fig.\ref{PhiPDMS}.a, we observed two types of response, depending on the volume fraction of polymer.  For $W$ between $0.3$ and $2.7$\%, $E$ depends strongly on $W$, and varies by more than an order of magnitude over this range.  Above $W= 2.7\%$, $E$ continues to increase with added polymer, but much more slowly than below the threshold value.    

By looking at pieces of these samples through a microscope we found that for volume fractions higher than $W= 0.7\%$ the polymer was homogeneously distributed throughout the sample, forming well-defined bridges between adjacent beads.  The samples were then ground up in a mortar, with a pestle, and a single layer of detached beads was spread on a glass slide, for each sample.  Images of these beads were taken by digital microscopy, the radii $r$ of the polymer bridges remaining adhered to the particles were measured manually using ImageJ \cite{Schneider2012}, and the bridge areas were calculated, assuming circular cross-sections.  At least 31 different bridges were analysed from each sample.   As shown in Fig.\ref{PhiPDMS}.b, the mean cross-sectional area of the bridges, $A$, increases linearly with $W$, up to several percent of polymer content. Trimers were first observed at $W^* = 2.7\%$, and this critical value is shown on Fig.\ref{PhiPDMS}.a to correspond well to the change in elastic response of the cohesive granular media.  There is a similar kink at $W^*$ in the dependence of the packing fraction of the beads, $\phi_b$, with the amount of polymer (see Fig.\ref{PhiPDMS}.c).  Finally, trimers occur when three adjacent bridges touch each other, and this situation is expected when the bridge volume  $V_c\simeq 0.058 R^3$, with $R$ the radius of the beads, if one assumes that all the liquid goes into the bridges \cite{Scheel2008}.  For our 210 $\mu$m samples this predicts a critical bridge area of $A_c = 6.7\times 10^{3} \ \mu$m$^2$.  As Fig.\ref{PhiPDMS}.b shows, in this case $W^{*}$ also corresponds to the point where the upper tail of the distribution of $A$ crosses $A_c$, but where the average $A$ is still only around $4\times 10^{3} \ \mu$m$^2$. For higher polymer content, the largest bridges will coalesce, increasing the fraction of clusters in the samples. For example, one can see on Fig.\ref{PhiPDMS}.b that the resulting trimming of the bridge size distribution leads to a significant deviation from the linear relation between $A$ and $W$ for $W=4.6\%$.  Although we were not able to quantify the absolute fraction of trimers, or larger clusters, these results qualitatively agree with the behaviour of wet granular packings, which suggests a ratio of trimers to bridges of about 5\%, when $W=3.5\%$ \cite{Scheel2008,Herminghaus2013}. 
 
These observations all show how the stiffness of an aggregate sample is related to the sizes and shapes of its bridges. As long as the bridges remain isolated, then $E\sim A$, as in Fig.\ref{PhiPDMS}.d.  For the lowest $W$ the polymer does not spread smoothly through the sample, and the result is a softer sample than otherwise expected.  If $W>W^*$, then any further addition of polymer will end up mainly filling the pore space between groups of particles, rather than strengthening the elastic bonds that hold the sample together.  This suggests that the slow increase in $E$ observed for $W>W^{*}$ could then result from the growth of bridges from the smaller end of their size distribution, rather than from pore filling and the coalescence of bridges into more complex structures.
 
\subsection*{Variation with volume fraction and bead diameter}

\begin{figure}
\centering
\includegraphics[width=1\linewidth]{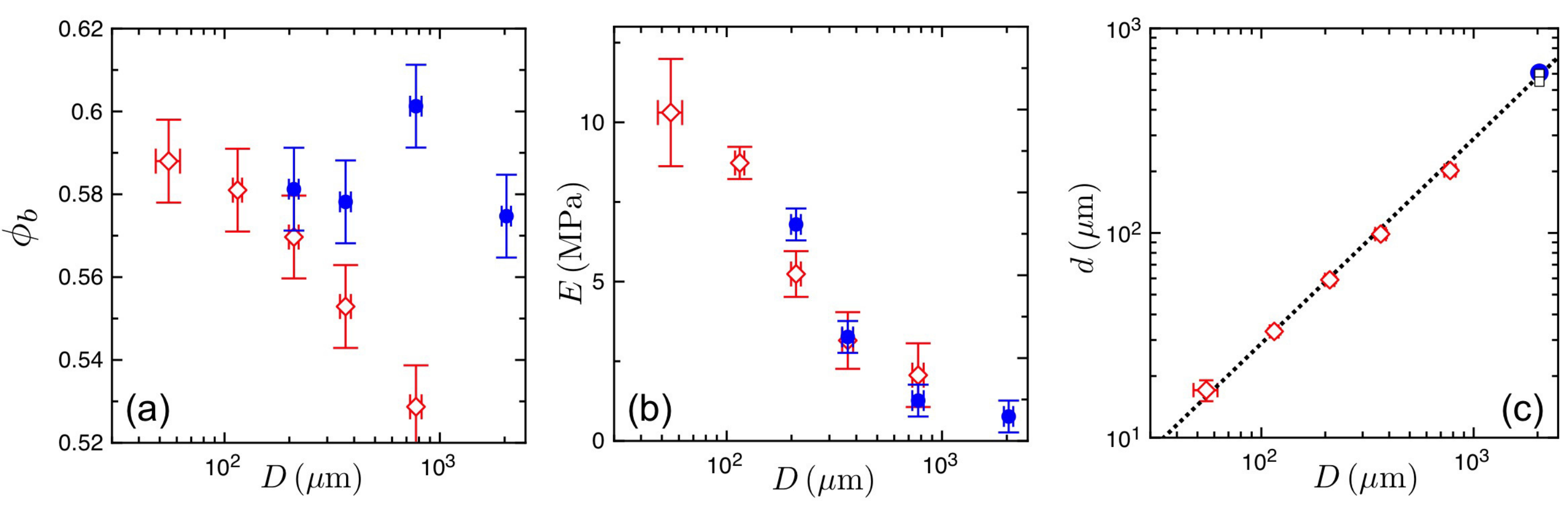}
\caption{The properties of the cohesive granular material depend on the packing fraction, $\phi_b$, and size, $D$, of its beads.  Samples were prepared at both a fixed aspect ratio (open red diamonds) and a larger sample size, but varied aspect ratios (filled blue points, see Methods for details).  Shown are (a) the average $\phi_b$ and (b) Young's modulus $E$, for these cases, as well as (c) the diameter $d$ of the bridges between particles. The open black dots in (c) at $D=2040 \ \mu m$ correspond to the drainage experiment described in text. In (c) the dashed line represents a linear relation between $D$ and $d$.
Error bars give standard errors.}
\label{Diameter}
\end{figure}

As we have seen, the volume fraction of the beads, $\phi_b$, in fully cured samples may change along with other parameters such as $E_p$ and $W$. This may affect the density of bead contacts, or bridges, in different samples.  It is therefore necessary to test for any dependance of $E$ on $\phi_b$.  To do so we used a coupling between $\phi_b$ and the diameter of the beads $D$ in finite size samples to disturb the homogeneity of the packing and vary the average value of $\phi_b$.  In particular, we found that confinement in a mould leads to an ordering of the outer layer of beads, as in other granular media \cite{Desmond2009, Jerkins2008}.  This creates a different packing of particles near the walls, than in the bulk.  By changing the size of such samples, relative to the size of their constituent beads, we could thus adjust their total average volume fraction.

In a first series of experiments we kept the geometry of the samples fixed to a cylinder of 4.65 mm $\times$ 4.9 mm (base diameter $\times$ height), while increasing the bead size from $D=55 \, \mu$m to $D=775 \, \mu$m.   We also  fixed $W=2.3$\% and $E_{P}=1$ MPa.   As shown in Fig.\ref{Diameter}.a, the mean packing fraction in these samples varied from $\phi_b = 0.59$ to 0.53, a range at least as broad as those already encountered in Figs. \ref{Young}.b and \ref{PhiPDMS}.c.  Next, by using moulds with dimensions much larger than the bead size (\textit{i.e.\,}at least 25 beads across, see Methods), we sought to minimise boundary effects, and keep $\phi_b$ constant.  As demonstrated in Fig.\ref{Diameter}.a, the volume fractions of these larger samples were all consistent with an average value of $\phi_b = 0.584\pm0.01$, regardless of particle diameter.   These two sets of experiments, summarised in Fig.\ref{Diameter}.b, allowed us to independently test the effects of the beads' packing density and size on the elasticity of the cohesive material. We found no significant dependance of $E$ on $\phi_b$, but instead observed an unexpected variation in $E$, of nearly one order of magnitude, across the range of bead sizes tested.
 
This dependence of material elasticity on bead size is somewhat surprising, as linear elasticity is inherently scale-free.  In other words, if all lengths in the system are scaled up by a constant factor, we would expect the elastic response to remain unchanged.  That this is not, in fact, true here suggests that there is an additional length-scale in the system, such as a minimum bead separation.  One length-scale that can be excluded from consideration here is the capillary length.  For the samples discussed above we also measured the average bridge and bead radii.  As shown in Fig.\ref{Diameter}.c, there is no deviation from the linear relation between $d$ and $D$ expected for capillary bridges \cite{Willett2000, Herminghaus2005}, showing that gravity can be safely neglected here, even for millimetric beads. Any significant effect of drainage of the liquid polymer in the granular pile prior to curing was also rejected by measuring the size of the bridges in a cylindrical sample of 16 mm x 38 mm  (base diameter x height) made with the biggest (2 mm) beads. After curing the sample was cut into three pieces of 1cm height and the bridge size distribution was measured for each piece, showing almost no differences with respect to the position in the sample, as shown in Fig.\ref{Diameter}.c.

To summarise, we have shown that perturbations of the average volume fraction of the beads appear to have little observable influence on the mechanical properties of the material. However, we \textit{have} observed a strong dependance of $E$ on the bead size, which implies that there is some additional length-scale that is relevant to the physics here.  Although the origins of this effect remain unclear, we have ruled out a possible influence of drainage.  

\subsection*{Behaviour at larger compressive strains}

So far we have limited our study to the linear regime of the stress-strain curves, from which the Young's modulus of a sample can be extracted. Now, we briefly investigate the behaviour of the cohesive granular material for larger compressive strains, focusing on  the end of the linear regime.  Although deviation from a linear elastic response is not necessarily equivalent to yielding of the sample, we shall presently show how it corresponds to the onset of shear failure in the compressed sample.  We therefore take it to be a good approximation of the yielding point.   

\begin{figure}[h!]
\centering
\includegraphics[width=1\linewidth]{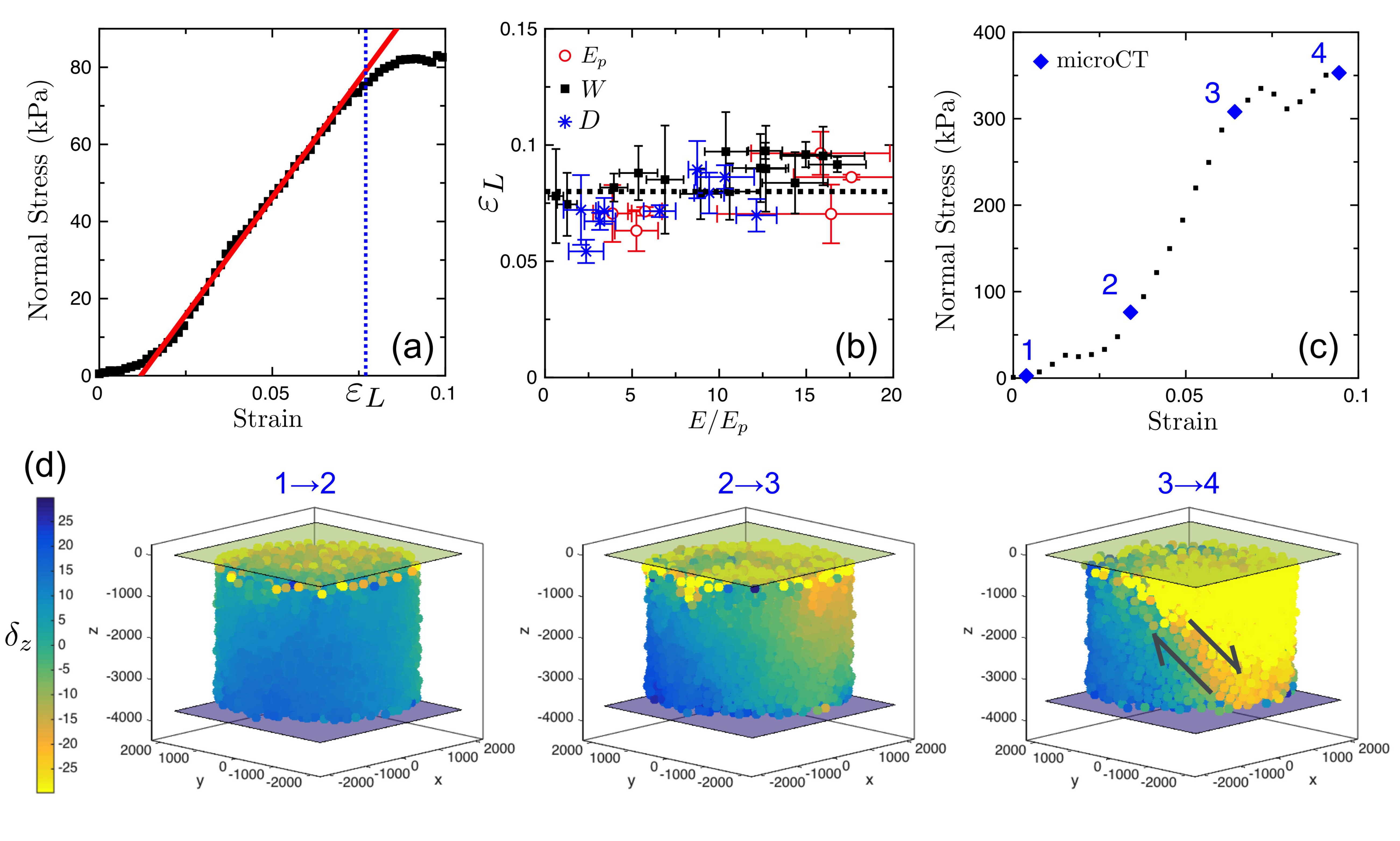}
\caption{(a) Determination of the limit strain $\epsilon_L$ for a typical stress-strain curve. Red line: best fit of the linear regime, giving $E=1.20\pm0.05$ MPa. $\epsilon_L$ is defined as the first point significantly deviating from the linear relation, and is here $\epsilon_L=7.7\pm0.2\%$.    (b)  $\epsilon_L$ as a function of relative Young's modulus of the cohesive granular material, $E/E_P$, for all experiments (open circles: varying $E_P$, black squares: varying $W$, blue stars: varying $D$). Dashed line: average value of 8\% limiting strain. (c) Stress-strain curve of the sample used for the \textit{in situ} compression test. Each blue diamond corresponds to an x-ray micro-computed tomography scan.  (d) Displacement of each bead along the z-axis, the direction of compression, between scans 1 and 2 (left), scans 2 and 3 (middle), and scans 3 and 4 (right). The colour bar indicates the magnitude of $\delta_{z}$ in microns. The top and bottom planes show the positions of the pistons, in microns. The arrows indicate the developing shear failure of the sample. }
\label{EndLinear}
\end{figure}

For each of the compression experiments discussed above we determined the limit strain $\epsilon_L$ by the first point of the stress-strain curve that deviates significantly from a linear fit, as demonstrated in Fig.\ref{EndLinear}.a.  The results are shown in Fig.\ref{EndLinear}.b.  Interestingly, we found that all samples, whether they varied in bead size, bridge size, or bridge composition, were consistent with a constant $\epsilon_L=$ 8\% (positive one-tailed t-test with a p-value of 0.49).  This suggests that the plasticity of the material is controlled by a limiting \textit{strain}, rather than a limiting stress.  

To explore further the internal deformations of the sample, in this limit of large strains, we performed an \textit{in situ} uniaxial compression experiment on one of our samples using micro-computed x-ray tomography (microCT); details of this test are given in the Methods section.   The use of precision monodisperse beads here allowed us to reconstruct the positions of nearly all the beads in the sample after a series of compressive steps.  Four x-ray scans were made at different strains, as indicated in Fig.\ref{EndLinear}.c, and the position vectors $\vec{x}_{i,n}$ were found for each bead $i$ after each compression step $n$.  The relative displacements of the beads, $\vec\delta_{i,n}$, were calculated as $\vec\delta_{i,n}= \vec x_{i,n}-\vec x_{i,n-1} - \langle \vec x_{i,n}-\vec x_{i,n-1} \rangle$, where the angular brackets represent averaging over all beads (\textit{i.e.\,}we subtract the sample-wide average displacement).

The relative displacements within the sample are visualised in Fig.\ref{EndLinear}d.    There we show how the displacement field evolves toward a non-linear response during the compression test.  Analysis of the first 2$\%$ of the strain (between points 1 and 2) confirms that this initial phase corresponds to the progressive contact of the rough surface of the sample with the piston.  Variations of the displacement field in the linear regime between points 2 and 3 remain smooth, as expected for a linear elastic response, other than a few isolated features (slip events, local rearrangements) near the edges of the load cell. However, it is interesting to note that the displacement field is not entirely uniform and already shows the onset of a gradient that will evolve towards a shear band in the plastic regime.  For higher strains, between points 3 and 4, we cross the limiting strain $\epsilon_L$.  Now we can clearly see the intermediate stages of shear failure, demonstrated by the shear band, or slip plane, that has developed across the sample.  This type of yielding behaviour is expected for homogeneous and isotropic solids failing under compression, when the deviatoric stress (or strain) reaches the yield stress (or strain). Several amorphous materials, such as metallic glasses \cite{Johnson1999}, granular materials \cite{Li2015}, or foams \cite{Kabla2007}, also form shear bands when failing. However, the exact process by which these local plastic rearrangements lead to final persistent shear planes at failure remains unclear and is still a topic of active research \cite{Maloney2006, Lebouil2014}.

\section*{Discussion}

We have shown how to make stiff, porous granular materials, and characterised their mechanical properties.  As demonstrated above, the elastic response of this class of cohesive material can be tuned by changing the stiffness and size of its polymer bonds, as well as the diameter of its beads.  A full micro-mechanical model, which would link macroscopic responses such as elasticity, fracture and plasticity to microscopic modes of deformations, is beyond the scope of this paper. Nevertheless, we will see here how simple models of elastic bridges, combined with finite element simulations (performed using Comsol Multiphysics), can be used to derive simple scaling relations that can explain most of the observed findings.

In these materials there is a large difference between the Young's moduli of the glass beads and of the polymer bonds: $E_p/E_b<10^{-4}$.  Since the modulus of the composite material also scales as $E\sim E_p$, this suggests that we can neglect the deformation of the beads when considering elastic responses.  Indeed, the stress transmitted by actual contact of the beads should obey the Hertz theory of contacts \cite{Hertz1881}, and lead to a strongly non-linear constitutive relation \cite{Makse2004}.  This is not what we see.  

A linear elastic response can also be observed when granular materials are cemented with a hard cement, when $E_p \simeq E_b$.  This is the case for glass or PMMA cemented by epoxy \cite{Dvorkin1999} or frozen tetradecane \cite{Langlois2014}, for example. However, in such systems force propagation still occurs by Hertzian contacts between the beads, in parallel with deformation of the cement \cite{Dvorkin1999,Langlois2014}. This leads to a stiffness of the packing that is comparable to that of the grains: $E\sim E_b$.  Again, this is not what we see: here $E/E_b<10^{-3}$. 

\begin{figure}
\centering
\includegraphics[width=1\linewidth]{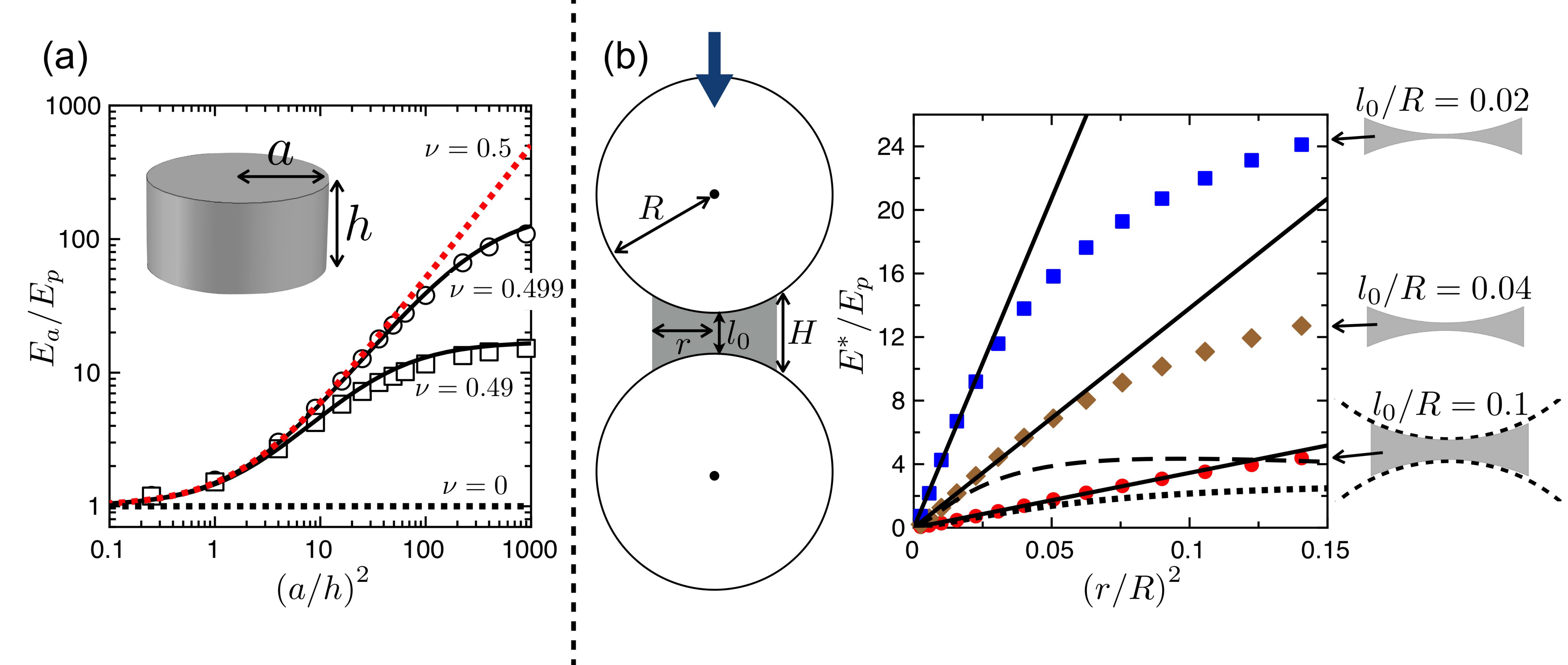}
\caption{(a) The apparent Young's modulus $E_a$ of a cylinder, compressed between two plates \textit{without} slip, can be much higher then that of the material composing it, $E_p$, and is a function of the aspect ratio $a/h$ and Poisson ratio $\nu$. The curves give solutions\cite{Williams2008} of Eq. \ref{Williams} for various $\nu$, while the symbols show the results of corresponding numerical simulations. (b) Left: sketch of a model geometry of two hard spheres connected by a compliant bridge under uniaxial compression. Right: normalised apparent Young's modulus $E^*/E_p$ as a function of relative bridge size $(r/R)^2$. Dots correspond to simulations at $\nu = 0.49$ and three different gap heights. Broken lines give solutions of Eq. \ref{WilliamsStar} for $l_0/R=0.1$ and 0.04. Solid lines: $E^*/E_p\propto (r/R)^2$.}
\label{Figure_Comsol}
\end{figure}

Therefore, we will instead consider a model where the elastic response of the aggregate is only due to elastic deformations of the polymer bridges. Interestingly, the effective Young's modulus of such a model can still be an order of magnitude higher than that of the bridge material, a result similar to the data shown in Fig.\ref{Young}.c. This can be understood by first considering the simple example of the uniaxial compression of a cylinder of radius $a$ and height $h$ with bonded surfaces (\textit{i.e.\,}there is no slip between the sample and the pistons, during compression). Williams and Gamonpilas \cite{Williams2008} derived an analytic relation between the ``apparent'' measured Young's modulus of this cylinder, $E_a$, and the true Young's modulus of the material composing it, $E_p$:
\begin{equation}
\frac{E_a}{E_p}=\frac{1+3\nu\left (\frac{1-\nu}{1+\nu}\right )S^2}{1+3\nu(1-2\nu)S^2},
\label{Williams}
\end{equation}
where $\nu$ is the Poisson ratio of the material and  $S=a/h$ the aspect ratio of the cylinder.  We show the results of Eq. \ref{Williams} in Fig.\ref{Figure_Comsol}.a. and compare them with a finite element simulation of a bonded cylinder under uniaxial compression. Here $E_a$ can be significantly higher than $E_p$, especially for large aspect ratios and for materials with a Poisson ratio close to 0.5, such as is the case for elastomers. For perfectly incompressible materials ($\nu=0.5$), Eq. \ref{Williams} reduces to $E_a/E_p = 1+ S^2/2$, and for large aspect ratios, $S\gg1$, it further shows a scaling of the apparent stiffness with the cross-sectional area of the cylinder. Although a cylinder is only a rough approximation of a solid capillary bridge, this example shows how bonding between the beads and the bridges, along with incompressibility of the polymer bridges and their large aspect ratios, can lead to a Young's modulus of the aggregate that is higher than that of the polymer. Also, the model predicts a linear relation between $E_a$ and $E_p$, and a scaling of $E_a$ with the cross-sectional areas of the polymer bridges, in fair agreement with the measurements reported in Figs. \ref{Young}.c and \ref{PhiPDMS}.d, respectively.

To go towards a more realistic description of the mechanics of the polymer bridges, we consider now a system made ip of two beads of radius $R$, initially separated by a surface-to-surface distance $l_0$ and connected by an elastic bridge consisting of a cylinder of radius $r$, height $H$ and Poisson ratio $\nu$, which has been truncated by the spherical caps of the intruding beads (see Fig.\ref{Figure_Comsol}.b). As in the example discussed above, the bridge is firmly bonded to the beads, so that its contact areas remain fixed. We performed a series of simulations in this configuration by displacing the centres of both beads towards each other, and measuring the total force required to hold them there (see Methods).  The effective strain, $\epsilon^*$, was defined as the relative change of the centre-to-centre distance between the beads, and the effective stress, $\sigma^*$, as the restoring force divided by the bead cross-sectional area.   In the limit of small strains the relationship between $\epsilon^*$ and $\sigma^*$ is linear and does not depend on the sign of the deformation (\textit{i.e.\,}compression vs. tension).  We thus define an effective Young's modulus as $E^*=\sigma^*/\epsilon^*$, for small strains.   The results of these simulations are shown in Fig.\ref{Figure_Comsol}.b for different gaps $l_0$ and bridge sizes $r$, and taking nearly incompressible bridges with $\nu=0.49$.  As in the case of the bonded cylinder, we see that here $E^*$ can be significantly higher than $E_p$, in particular for beads close to contact (\textit{e.g.\,}for $l_0/R=0.02$).   Specifically, like the data shown in Fig.\ref{Young}.c, the bridge-particle assembly is up to about an order of magnitude stiffer than the material in the bridge itself.

These simulations allow us to test the range of the analytic model that treats an elastic bridge as a simple bonded cylinder.  To do so, we replace the bridge between the beads by a full (\textit{i.e.\,}untruncated) cylinder of radius $r$ and height $H$ (see Methods and Fig. \ref{Figure_Comsol}.b).
We see in Fig.\ref{Figure_Comsol}.b that this approximation is valid when the bridge has a shape close to a cylinder ($H \sim l_0$, \textit{i.e.\,}for small $r/R$ or large $l_0/R$) but that it fails to describe the cases of large bridges and beads close to contact. Over a larger range of parameter space the simulations instead match the relation $E^*/E_p \propto (r/R)^2$, which is in good agreement with our measurements of $E\propto r^2$ in the pendular regime. Figure \ref{PhiPDMS}.d showed that this result holds experimentally up to $(r/R)^2\simeq0.12$.   It is known that in wet granular assemblies, capillary bridges are formed between beads in contact or very close to contact \cite{Herminghaus2005}. However, liquid bridges can also span a finite distance, so that the average number of capillary bridges per bead is at least $\sim 10-15$\% higher than the number of dry contacts for a given packing fraction \cite{Herminghaus2005,Fournier2005}. The wide range of validity for the relation $E\propto r^2$ in our experimental results could then indicate that the elastic behaviour of the aggregate is mainly governed by these extended bridges.

We have seen that simple contact-mechanics models can enlighten the underlying physics of our system and provide good quantitative comparisons to our experimental findings.  They even suggest the importance of small gaps between beads, when bridges are made, to the macroscopic response of the material.  An additional characteristic length scale, such as this gap width, is necessary to explain the observed dependence of $E$ on the bead size (Fig.\ref{Diameter}.b).  However, it also shows that other small-scale features could affect the scaling of the elastic response of cohesive granular materials.  To explore models of these materials further would require a detailed micro-mechanical understanding of other modes of deformation, besides tension or compression,  such as shear, torsion or bending.    More realistic numerical simulations of the whole packing, similar to those describing the behaviour of aggregates cemented with stiff cements \cite{Holtzman2012a, Affes2012, Carmona2008}, could also clarify, for example, the role of intergranular friction and bead roughness. Furthermore, the link between the microscopic details (\textit{i.e.\,}at the bridge or bond-scale) to the macroscopic elastic properties of the sample could be investigated using lattice-solid models, which simulate large ensembles of discrete particles interacting via springs, and which are widely used to model the mechanical properties of rocks \cite{Place1999, Wang2006, Zhao2010}. However, one of the main differences between existing models and our system is the peculiar dependance of the bridge stiffness on the initial bead separation, which originates principally from the incompressibility of the polymer bonds.

In several places we have noted that our materials can behave differently to the more well-known cemented aggregates.   One useful feature of our cohesive aggregates worth highlighting is the large strains that they can accommodate. The value of the limiting strain is $\epsilon_L\sim 8$\%, compared to typical values of $1-2$\% of yield strain observed for aggregates cemented with hard cement\cite{Delenne2009}.  Elastomers such as PDMS remain elastic for very large deformations (\textit{e.g.\,}PDMS can be reversibly stretched by more than 40 \% of its length\cite{Johnston2014}), perhaps explaining this difference. The fact that yielding is observed for a constant strain, largely independent of sample stiffness and of the microscopic parameters $W$, $E_p$ and $D$ is remarkable. This behaviour indicates that yielding of the aggregate could be a purely geometric effect linked to local rearrangement of the beads, in a similar manner to the yielding of amorphous systems such as colloidal suspensions \cite{Boulogne2014} and dense emulsions \cite{Hebraud1997}, for example. Further understanding of the yielding properties of the material via modelling would require a micromechanical model taking into account plastic deformations at the bridge scale, for example partial debonding or fracture at the contacts \cite{Leopoldes2013}, as observed in failure of weakly cemented granular materials \cite{Langlois2014}.

In summary, we have investigated the mechanical properties of a cohesive granular material obtained by mixing glass beads with a curable elastomer, and shown how its elasticity can be finely tuned by controlling the stiffness of the polymer phase and its volume fraction within the sample. We have seen how scaling relations for the Young's modulus of the aggregate can be obtained by using simple models, which in turn show good agreement with our experimental findings. We have also found that the stiffness of the aggregate depends on the bead size, while remaining largely insensitive to variations in the volume fraction of the beads. Finally, we have shown that this material exhibits a linear elastic response, even for large strains, and a strain at yield that is independent of its strength and microscopic details. The high tunability of the properties of this model cohesive granular material makes it a good candidate for investigations on problems involving cohesive porous media, such as the hydraulic fracture of porous rocks, seismology, bio-fouling, or root growth in porous matrices. 

\section*{Methods}
\subsection*{Glass beads}

All beads were made of soda-lime glass with a Young's modulus between 60 GPa and 70 GPa (manufacturer's info). Prior to use, beads were first cleaned for two hours in a strong surfactant solution (20\%, by weight, Hellmanex III, Hellma Analytics, diluted with water), rinsed, further cleaned with a solution of 1M NaOH, and rinsed again several times with water, until reaching neutral pH.  Deionized (Millipore) water was used for all cleaning steps.  Finally, the beads were dried in an oven at 90$^{\circ}$C overnight.

The densities of each type of bead were measured with a precision of $\pm5$ kg/m$^{3}$ using a pycnometer, and the results are summarised in Table \ref{tab:beads}.  The 210 $\mu$m beads were used for most tests, while the monodisperse 200.9 $\mu$m beads were reserved for the tomography experiments.  The remaining beads were used to test the effects of particle size on material properties.   The particle size distributions of the beads were measured by optical microscopy using a minimum of 100 beads per bead type.  The mean diameters agreed with manufacturer specifications to better than 5\%, and are shown in Table \ref{tab:beads}.  The polydispersity, also shown, corresponds to the standard deviation of the distribution of bead diameters, divided by the mean.  To limit the effects of packing effects near the walls of the container, during casting and curing, the diameter and height of the samples could be varied along with bead size.  The various specimen sizes are also listed in Table \ref{tab:beads}.

\begin{table}[ht]
\centering
\begin{tabular}{|l|l|l|l|l|l|}
\hline
Mean diameter ($\mu$m) & Polydispersity & Density (kg/m$^{3}$) & Producer & Diameter (mm) & Height (mm) \\
\hline
55  & 15\% &2460 & Sigmund Lindner & 4.65 & 4.90 \\
\hline
115  & 5\% &2485  &Whitehouse Scientific & 4.65 & 4.90\\
\hline
210  & 5\% &2495 & Sigmund Lindner  & 4.65/ 12.10 / 15.00 & 4.90 / 7.00 / 7.70\\
\hline
365  & 5\% &2495 &Whitehouse Scientific & 4.65 / 22.05 & 4.90 / 11.35 \\
\hline
775  & 5\% &2570 &Whitehouse Scientific & 4.65 / 50.00  & 4.90 / 25.10    \\
\hline
2040  & 5\% &2565 & Sigmund Lindner & 4.65 / 50.00 & 4.90 / 26.50  \\
\hline
\hline
200.9  & $<$1\% &2500 &Whitehouse Scientific  & 4.25 & 4.05 \\
\hline
\end{tabular}
\caption{\label{tab:beads} Properties of the beads and of the specimens.}
\end{table}

\subsection*{Sample preparation}
PDMS was prepared by thoroughly mixing Sylgard 184 (Dow Corning) base and curing agent in the desired weight ratio (from 10:1 to 66:1) and then degassing the resulting mixture under vacuum. The density of the degassed PDMS was measured to be 1010 kg/m$^{3}$, and did not vary within instrument precision over the range of mixing ratios used.  Samples were made by mixing a set amount of PDMS uniformly into glass beads (typically $\sim 10$ mL of beads) in a mortar with a pestle, taking care that the PDMS was imbibed directly into the bead pack, rather than adhering to the pestle or mortar wall. A subset of the mixture was then cast into a desired mould, where it was gently compressed into shape. To ensure sample homogeneity, the moulded samples were vibrated at a fixed acceleration of $2g$ for two minutes at each of the following frequencies: 20 Hz, 100 Hz, 200 Hz, and 500 Hz. This provided an acceleration that is close to, but below, the fluidisation onset of wet granulates \cite{Fournier2005}. After resting for one hour, to ensure equilibration of the capillary bridges, samples were then baked at 75$^{\circ}$C for 14 hours.  The volume fraction of the beads, $\phi_b$, in any sample was obtained by measuring the mass ($m_s$) and volume ($V_s$) of that sample, the density of the beads $\rho_b$, and the masses of PDMS ($m_p$) and beads ($m_b$) mixed in the mortar prior to curing, and using the relation $\phi_b= \rho_s m_b/ \rho_b (m_p+m_b) $, where $\rho_s = m_s/V_s$.

\subsection*{Mechanical tests}
Unconfined uniaxial compression tests were performed using two different custom-made testing machines.  The results from the two instruments showed no noticeable differences, for tests on samples prepared under identical conditions.  The Young's modulus of PDMS, $E_P$, was determined from cylindrical samples of dimensions (base diameter $\times$ height) of 15.00 mm $\times$ 8.00 mm,  while the Young's modulus of the cured aggregate samples, $E$, was measured on cylinders of dimensions 4.65 mm $\times$ 4.90 mm, unless stated otherwise. The cylinder heights could vary by $\sim$10\% between samples, but were measured individually for each test, and the measured values were used in calculating the sample strain.  A droplet of sunflower oil was deposited on the surface of the pistons before each experiment, to prevent sticking of the sample during compression. The results do not depend on the sample size or aspect ratio, as one can see on Figure \ref{Young}.c., where we show results of three measurements at different $E_p$ for cylinders of diameter 15.00 mm x height 7.70 mm, at W=2.3\%, D=210 $\mu m$ and $\phi_b=0.56-0.57$.

During a mechanical test compression was controlled \textit{via} a stepper motor of submicron resolution, while the restoring force was monitored using either an analytical balance (Denver Instruments) or a compact high-precision load cell (model 31E, Honeywell), with an effective force resolution of 5 mN.  Compression consisted of discrete steps of 7.6 $\mu$m.  After each step the load cell was continuously monitored, until the restoring force reported by it had reached an equilibrium value.  Only then was the next compression step made.  This led to an average compression rate $\dot\gamma$ of about 5 $\mu$m/min.  Tests made with significantly longer ($\dot\gamma \simeq 0.8 \ \mu$m/min) or shorter ($\dot\gamma \simeq 20 \ \mu$m/min) intervals between compression steps showed no significant variation in the resulting stress-strain curves.  Each value of $E$ given here is the result of at least three independent measurements, with standard deviations shown as error bars. 

\subsection*{X-ray microtomography}
The sample used for \textit{in situ} x-ray microtomography was prepared with monodisperse beads of diameter $200.9  \pm 1.9 \, \mu$m (see Table \ref{tab:beads}) using a volume fraction $W=3.7$ \% of PDMS prepared with a 10:1 weight ratio of base to curing agent ($E_P=1.5$ MPa). The Young's modulus of the sample was measured as $E=7.9$ MPa and the bead volume fraction was $\phi_b=0.59\pm0.01$.   A series of four x-ray computed microtomography scans (GE Nanotom) was performed at different fixed strains (see Figure \ref{EndLinear}), using the setup sketched in Figure \ref{Beads}.d.   The sample was slowly compressed ($\dot\gamma\simeq5 \mu$m/min) by 120 $\mu$m between each scan, using the same methods described above for mechanical testing.  Tomograms were acquired with a tungsten target and an acceleration voltage of 120 keV. Each scan consisted of a set of 2000 projections with a resolution of 1132$\times$1132 pixels and a voxel size of 5 $\mu$m. Beads were detected within the reconstructed volume using Matlab. In particular, after thresholding using Otsu's algorithm \cite{Otsu1979} and 3D erosion of the binary images to separate neighbouring beads, particle positions were determined by finding the centroids of each individual connected volume.  By using beads that were monodisperse, both in size and shape, we could detect all beads within the sample and measure their centre-positions to within about one voxel's precision.  The displacements in bead positions between subsequent scans were generally small, as compared to the bead diameter.  We thus tracked the motion of individual beads through the compression by assuming that the two closest bead positions, from any pair of successive scans, correspond to the same particle.

\subsection*{Comsol modelling}
Comsol Multiphysics 4.0 was used to build a 3-D finite-element model of an elastic bridge. The bridge was generated by subtracting two spheres of radius $R$ and surface-to-surface separation $l_0$ from a cylinder of Young's modulus $E_p$, Poisson's ratio $\nu=0.49$ and radius $r$ (see Fig.\ref{Figure_Comsol}.b). A finite translation $\delta$ was imposed on one of the sphere-bridge interfaces while the other one was kept fixed, using no-slip boundary conditions for both. The total reaction force $F_n$ exerted on the displaced interface was measured, giving an effective Young's modulus $E^*=F_n (2 R+L_0)/(\delta \pi R^2)$.\\
We also performed simulations of full cylinders to compare with analytical results obtained by Williams and Gamonpilas \cite{Williams2008} (see Fig.\ref{Figure_Comsol}.a), and to test modelling of a bridge with such a bonded cylinder of height $H=l_0+2R(1-\sqrt{1-r^2/R^2})$ (see Fig.\ref{Figure_Comsol}.b). Adapting Eq. \ref{Williams} to this geometry and our definition of $E^*$ leads to: 
\begin{equation}
\frac{E^*}{E_p}=\left ( \frac{1+3\nu\left (\frac{1-\nu}{1+\nu}\right )(r/H)^2}{1+3\nu(1-2\nu) (r/H)^2} \right ) \left ( 1 + 2 \frac{R}{H}\sqrt{1-\frac{r^2}{ R^2}} \right ) \frac{r^2}{ R^2}.
\label{WilliamsStar}
\end{equation}

\bibliography{biblio_goe.bib}

\section*{Acknowledgements}
The authors would like to thank W. Keiderling and M. Richter for technical assistance and help designing the compression tests, and R. Holtzman, C. MacMinn, and S. Herminghaus for fruitful discussions.

\section*{Author contributions statement}

A.H., M.S. and L.G. conceived the experiments.  A.H. conducted the experiments and analysed the results. A.H. and L. G. wrote the manuscript, while all authors discussed the results and reviewed the manuscript. 

\section*{Additional information}

\textbf{Competing financial interests:} A.H., M.S. and L.G. are inventors on a patent application regarding the cohesive granular material (PCT/EP2016/067079).

\end{document}